\documentclass[12pt]{iopart}
\usepackage{graphicx}
\usepackage{epstopdf}  
\begin{document}
\title[Thermoelectric energy converters under a trade-off figure of merit with broken time-reversal symmetry]{Thermoelectric energy converters under a trade-off figure of merit with broken time-reversal symmetry}
\author{I. Iyyappan and M. Ponmurugan}
\address{Department of Physics, School of Basic and Applied Sciences, Central
University of Tamil Nadu, Thiruvarur 610 005, Tamil Nadu, India.}
\ead{iyyap.si@gmail.com and ponphy@cutn.ac.in}
\vspace{10pt}
\begin{abstract}
We study the performance of a three-terminal thermoelectric device such as heat engine and refrigerator with broken time-reversal symmetry by applying the unified trade-off figure of merit ($\dot{\Omega}$ criterion) which accounts for both useful energy and losses. For heat engine, we find that a thermoelectric device working under the maximum $\dot{\Omega}$ criterion gives a significantly better performance than a device working at maximum power output. Within the framework of linear irreversible thermodynamics such a direct comparison is not possible for refrigerators, however, our study indicates that, for refrigerator, the maximum cooling load gives a better performance than the maximum $\dot{\Omega}$ criterion for a larger asymmetry. Our results can be useful to choose a suitable optimization criterion for operating a real thermoelectric device with broken time-reversal symmetry.
\end{abstract}
\section{Introduction}
	The increase of energy demands and rapid depletion of non-renewable energy resources calls for a renewable and eco-friendly energy resources. A thermoelectric device is a suitable candidate to meet these demands as a thermoelectric heat engine converts waste heat into electric power and a thermoelectric refrigerator converts electrical power into cooling directly. The upper bound of energy conversion is given by the Carnot efficiency, $\eta_C=1-T_c/T_h$ for heat engines and the Carnot Coefficient Of Performance (COP), $\eta^{r}_C=T_c/(T_h-T_c)$ for refrigerators, where $T_h$ and $T_c$ are the temperatures of the hot and cold reservoirs, respectively. To achieve the Carnot performance, the heat engines (refrigerators) need to operate reversibly which takes an infinite amount of time, i.e., with the vanishing power output (cooling power). However, any real heat devices have to work at finite power output (cooling power).
	
	Curzon and Ahlborn (CA) studied the finite time Carnot heat engine. Where the engine absorbs (ejects) the heat during the isothermal expansion (compression) process while having a finite time contact with the heat reservoir. The input (output) heat is obtained by the linear Fourier heat transfer law and assuming the entropy production in finite time adiabatic processes is zero (i.e., so called the endo-reversible approximation), they found the efficiency at maximum power output as \cite{cur}
\begin{equation}\label{1b}
\eta_{CA}=1-\sqrt{\frac{T_c}{T_h}}.
\end{equation}
This result is also obtained for a thermoelectric generator \cite{agr}. A thermoelectric device were widely studied in recent years within the linear response regime \cite{nak,ben,sai1,san1,bra1,ben2,bra2,bal,ken,whi,ben3,bra3,maz1,pro} by applying the linear irreversible thermodynamics (LIT) \cite{ons,cal,kon}, where the temperature difference between the hot and cold reservoirs $\Delta T=T_h-T_c$ is assumed to be small compared to the reference value $T_h\approx T_c \approx T$. For systems with time-reversal symmetry, the maximum efficiency and the efficiency at maximum power output are governed by a single parameter called the figure of merit, which is given by $ZT=(\sigma S^{2}/\kappa)T$, where $S$ is the Seebeck coefficient, $\sigma$ is the electrical conductivity and $\kappa$ is the thermal conductivity. The maximum efficiency of a thermoelectric heat engine reads as \cite{ben}
\begin{equation}\label{1f}
\eta_{max}=\eta_C\frac{\sqrt{ZT+1}-1}{\sqrt{ZT+1}+1},
\end{equation}
and the efficiency at maximum power output as \cite{van}
\begin{equation}\label{1g}
\eta(P_{max})= \frac{\eta_{C}}{2}\frac{ZT}{ZT+2}.
\end{equation}
The $\eta_{max}$ and $\eta(P_{max})$, respectively, attains its maximum value $\eta_C$ and $\eta_{C}/2$, when $ZT\rightarrow \infty$. Experimental study showed that the most materials exhibit a very low $ZT$, which demands the search for highly efficient thermoelectric materials \cite{mah,maj,now,dre,yak,bel,casa,sny,vin,sha,bis,zhao,xie}. In the presence of magnetic field, the Carnot efficiency is attainable with the finite power output, when the constraints on the Onsager coefficients are imposed only from the positivity of the entropy production rate \cite{ben}, however, such a result is excluded when we consider current conservation for noninteracting system  \cite{bra1}. Theoretical and experimental studies of various models of two-terminal systems have failed to incorporate an asymmetry in  thermopower even in the presence of a perpendicular magnetic field $\textbf{B}$ \cite{sai1,san1,god,lang,wol,eom,jac}. The asymmetry $S(\textbf{B})\neq S(-\textbf{B})$ in thermopower can be achieved by adding noise by means of a third terminal (probe) in the noninteracting and also interacting system which induce an inelastic scattering in the system \cite{sai1,san1,ent,yam}, provided there are no net average particles and heat fluxes between the probe and the system.

	A unified optimization criterion: $\chi= zQ_{in}/\tau_{cyc}$ for any heat device is proposed by de Tom\'{a}s \textit{et al}. \cite{det}, where $\tau_{cyc}$ is the duration of cycle time. For heat engines (refrigerators) $z=\eta$ ($\eta^{r}$) and $Q_{in}=Q_h$ ($Q_c$) is the heat absorbed from the hot (cold) reservoir. Many real heat devices are unlikely to work at optimized $\chi$ criterion \cite{det,esp1}, it may work at trade-off between the power output (cooling power) and efficiency (coefficient of performance). Hern\'{a}ndez \textit{et al}. \cite{her} introduced a new optimization criterion denoted by $\Omega$, which accounts for both useful energy and losses. For heat engines $\Omega=(2\eta-\eta_{max})Q_h$, where $\eta$ and $\eta_{max}$ are the efficiency and the maximum efficiency, respectively. For refrigerators $\Omega=(2\eta^{r} -\eta^{r}_{max})W$, where $\eta^{r}$ and $\eta^{r}_{max}$ are, respectively, the coefficient of performance and the maximum coefficient of performance and $W$ is the work consumed by the refrigerators. The rate-dependent version of the $\Omega$ criterion is called as the target function, which is given for heat engines (HE) as
\begin{equation}\label{1c}
\dot{\Omega}_{HE} =2\dot{W}-\eta_{max} \dot{Q_{h}},
\end{equation}
and for refrigerators (RE) as
\begin{equation}\label{1d}
\dot{\Omega}_{RE} = 2\dot{Q_c}- \eta_{max}^{r}\dot{W},
\end{equation}	
where $\dot{Q_h}$ ($\dot{Q_c}$) is the input heat flux absorbed from the hot (cold) reservoir for heat engines (refrigerators) and $\dot{W}$ is the delivered (consumed) power by the heat engines (refrigerators). The dot denotes the quantity per unit time for steady-state heat devices or the quantity divided by one cycle time for cyclic heat devices. Several models of heat devices were studied under the maximum $\dot{\Omega}$ criterion \cite{sala,sala1,cis,cis1,det2,san,hu,lon,lon1,zha,aya} and found that the efficiency under the maximum $\dot{\Omega}$ criterion, $\eta(\dot{\Omega}_{max})$, lies between the maximum efficiency and the efficiency at maximum power output i.e., $\eta_{max}>\eta(\dot{\Omega}_{max})>\eta(P_{max})$ \cite{her}. 

	The efficiency of a linear irreversible heat engine working at maximum power is bounded below half of the Carnot efficiency and reaches  $\eta_C/2$ under the tight-coupling condition. In order to increase the performance of a heat engine, for practical applications, both the power and efficiency need to be as high as possible.  In this context, the  ecological criterion \cite{angu} which represents the best compromise between the power output and the entropy production rate of the environment has been used for the natural system \cite{angu2} and it is shown to increase the system efficiency \cite{angu,angu2,aria,long}. Since the ecological criterion depends on the local environment, it is not an easy task to use this as a suitable criterion for enhancing the system efficiency in the generalized framework \cite{her,aya}. However, it has been shown that the ecological criterion is equivalent to the $\dot{\Omega}$ criterion for a specific condition of maximum efficiency \cite{zha,aya}. Therefore, in the generalized framework of linear irreversible thermodynamics, we use the $\dot{\Omega}$ criterion for a better performance of a thermoelectric heat device. For more details about the usefulness of $\dot{\Omega}$ criterion can be seen in Refs. \cite{her,det2,san,lon,aya}.

 	This paper is organized as follows. In section 2, we briefly review the three-terminal system. In section 3, we study the efficiency under the maximum $\dot{\Omega}$ criterion and its bound on power. In section 4, we study the COP at maximum cooling load, the COP under the maximum $\dot{\Omega}$ criterion and its bound on cooling load. Finally, we conclude in section 5.
\section{Three-terminal System}
\begin{figure}[h]
\centering
\includegraphics[scale=0.58]{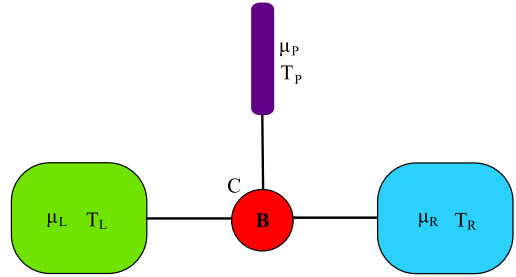}
	\caption{\label{fig:fig1} Schematic diagram of a thermoelectric device in the presence of a magnetic field $\textbf{B}$. The conductor C connected to three terminals of left and right reservoirs and a probe.}
\end{figure}

	The model system we consider in our study is shown in Fig. \ref{fig:fig1} \cite{sai1}. A system (conductor C) is in contact with the left (L) and  right (R) reservoirs. The temperature of the left and right reservoirs are $T_L=T+\Delta T$, $T_R=T$, respectively with $T_L>T_R$. Chemical potential of the left and right reservoirs are $\mu_L=\mu-\Delta\mu$ and $\mu_R=\mu$, respectively \cite{aper,fnote1}. Both the heat and electric currents flow along the horizontal axis. A third terminal (probe) reservoir at temperature $T_P=T+\Delta T_P$ and chemical potential $\mu_P=\mu+\Delta \mu_P$ is added to the system to induce inelastic scattering of an electron \cite{sai1}. The particle and energy currents of the $k$th reservoir  ($k=L, R, P$) are $J_{\rho k}$ and $J_{qk}$, respectively, flows into the system. From a steady-state constraints of charge and energy conservation, $\sum_k J_{\rho k}=0$ and $\sum_k J_{qk}=0$  \cite{sai1}. For heat engine the particle flows from the left to the right reservoir in which the particle current is taken as positive, therefore $\Delta\mu=\mu_R-\mu_L$ is positive. 
Since there is no net particle and heat flux between the probe and the system \cite{sai1}, the total entropy production rate of the reservoir can be written in the linear combination of fluxes ($\textbf{J}$) with the corresponding thermodynamic affinities ($\textbf{X}$) as 
\begin{equation}\label{1i}
\dot{S}=J_{\rho}X_{\rho}+J_qX_q,
\end{equation} 
where the thermodynamic affinities $X_{\rho} = -\Delta\mu/T$ drives the particle flux $J_{\rho}$ from the left to right
reservoir and $X_q = \Delta T/T^{2}$ drives the heat flux $J_q$ from the left to right reservoir. In linear response regime, the particle and heat fluxes can be written as \cite{sai1,ons}
\begin{equation}\label{1o}
J_{\rho}=L_{\rho\rho} X_{\rho} +  L_{\rho q}X_q,
\end{equation}
\begin{equation}\label{1o2}
 J_q = L_{q\rho}X_{\rho}+ L_{qq}X_q. 
\end{equation}
The Onsager coefficients $L_{ij}$ ($i,j=\rho,q$) satisfies the Onsager-Casimir relation
\begin{equation}\label{1p}
L_{i j}(\textbf{B}) = L_{j i}(-\textbf{B}).
\end{equation} 
The total entropy production rate becomes
\begin{eqnarray}\label{1q}
\dot{S}&=&J_{\rho} X_{\rho} + J_{q} X_q \\ \nonumber
&=& L_{\rho\rho}X_{\rho}^{2}+(L_{\rho q}+L_{q\rho})X_{\rho}X_q+L_{qq}X_q^{2}\geq0.
\end{eqnarray}
The positivity of the entropy production rate implies for the Onsager coefficients reads as \cite{ben}
\begin{eqnarray}\label{1r}
L_{\rho\rho} \geq0, \: L_{qq} \geq0, \: L_{\rho\rho} L_{qq}-\frac{1}{4}(L_{\rho q} +L_{q\rho})^{2}\geq0.
\end{eqnarray}
For noninteracting system, the additional constraint on the Onsager coefficients obtained from current conservation as \cite{bra1}
\begin{equation}\label{1u}
L_{\rho\rho}L_{qq}-(L_{\rho q}+L_{q\rho})^{2}/4\geq3(L_{\rho q}-L_{q\rho})^{2}/4.
\end{equation}
The above constraint gives rise to a stronger bound on the Onsager coefficients than the constraint obtained from the positivity of entropy production rate. For time-reversal symmetric case $L_{\rho q}=L_{q\rho}$, we recover  Eq. (\ref{1r}). The thermoelectric transport coefficients can be written in terms of the Onsager coefficients as \cite{cal}, namely the electric conductance $\sigma(\textbf{B})=e^{2} L_{\rho\rho}(\textbf{B})/T$, the thermal conductance $\kappa(\textbf{B})=\textbf{Det} \, \textbf{L}(\textbf{B})/ (T^{2} L_{\rho\rho}(\textbf{B}))$, the Seebeck coefficient $S(\textbf{B})=L_{\rho q}(\textbf{B})/( e T L_{\rho\rho}(\textbf{B}))$, and the Peltier coefficient $\Pi(\textbf{B})=L_{q\rho}(\textbf{B})/(e L_{\rho\rho}(\textbf{B}))$. Here $\textbf{Det} \, \textbf{L}\equiv L_{\rho\rho} L_{qq}-L_{\rho q}L_{q\rho}$. The thermopower becomes asymmetric when $L_{\rho q}(\textbf{B}) \neq L_{q\rho}(\textbf{B})$. Using the above relations, we can write the asymmetry parameter $x$ and the generalized figure of merit $y$ as \cite{ben} 
\begin{equation}\label{1z}
x\equiv\frac{L_{\rho q}\left(\textbf{B}\right)}{L_{q\rho}\left(\textbf{B}\right)}=\frac{S\left(\textbf{B}\right)}{S\left(-\textbf{B}\right)},
\end{equation}
\begin{equation}\label{1aa}
 y\equiv\frac{L_{\rho q}\left(\textbf{B}\right) L_{q\rho}\left(\textbf{B}\right)}{\textbf{Det} \, \textbf{L}(\textbf{B})}=\frac{\sigma\left(\textbf{B}\right) S\left(\textbf{B}\right) S\left(-\textbf{B}\right)}{\kappa\left(\textbf{B}\right)}T.
\end{equation}
We omit the magnetic field denoted by $\textbf{B}$ for simplicity in the rest of paper. Using Eqs. (\ref{1u}), (\ref{1z}) and (\ref{1aa}), it is identified as \cite{ben,bra1}
\begin{equation}\label{1ah}
h(x)\,\leq 4y \,\leq 0 \:~~ \mbox{if} \:~~ x<0,
\end{equation}	
\begin{equation}\label{1ai}
0 \,\leq 4y \,\leq h(x) \:~~ \mbox{if} \:~~ x>0, 
\end{equation}
where $h(x)= 4x/(x-1)^2$.
\section{A thermoelectric heat engine}
Once the system is reached a steady-state, the constant heat and particle flux flow through a conductor. The power output of heat engine is given by \cite{ben}
\begin{equation}\label{1ab}
P=J_{\rho}\Delta \mu= -T J_{\rho}  X_{\rho} 
\end{equation}
In a steady-state condition, the efficiency of a thermoelectric heat engine is defined as the ratio of the power output and the input heat flux extracted from the hot reservoir \cite{ben}, which is given by 
\begin{equation}\label{1ac}
\eta=\frac{-T J_{\rho}  X_{\rho}}{J_q}=\frac{-T(L_{\rho\rho}X_{\rho}^{2}+L_{\rho q}X_{\rho}X_q)}{L_{q\rho}X_{\rho}+L_{qq}X_q}.
\end{equation}
\subsection{The efficiency at maximum power output}
\begin{figure}[h]
\centering
\includegraphics[width=0.50\textwidth]{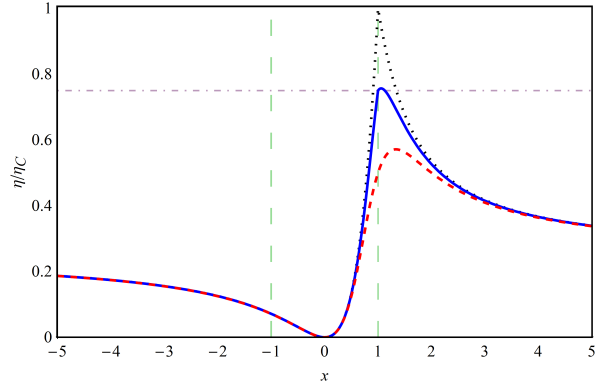}
\caption{\label{fig:fig2} The normalized efficiency $\eta/\eta_C$ is plotted as a function of the asymmetry parameter $x$. In the dotted curve $\eta=\eta_{max}$, the solid curve for $\eta=\eta(\dot{\Omega}_{max})$ and the dashed curve for $\eta=\eta(P_{max})$. The vertical dashed line represents $|x|=1$ and a horizontal dot-dashed line indicates $\eta=3\eta_C/4$.}
\end{figure}

To find the maximum efficiency, we maximize Eq. (\ref{1ac}) with respect to $X_{\rho}$ for fixed $X_q$, we get \cite{ben}
\begin{equation}\label{1ak}
X_{\rho}^{\eta_{max}}=\frac{L_{qq}}{L_{q\rho}}\left(-1+\sqrt{\frac{\textbf{Det} \, \textbf{L}}{L_{\rho\rho}L_{qq}}} \,\right)X_q,
\end{equation}
with the condition $J_q>0$. Substituting Eq. (\ref{1ak}) in Eq. (\ref{1ac}), we get the maximum efficiency as \cite{ben}
\begin{equation}\label{1al}
\eta_{max}= \eta_C x \frac{\sqrt{y+1}-1}{\sqrt{y+1}+1},
\end{equation}
where the Carnot efficiency, $\eta_C= T X_q$. For a given value of the asymmetry parameter $x$, $\eta_{max}$ attains its maximum  when $4y=h(x)$, which is given by \cite{bra2}
\begin{equation}\label{1am}
\eta_{max}=\eta_C x\frac{\sqrt{x^{2}-x+1}-|x-1|}{\sqrt{x^{2}-x+1}+|x-1|}.
\end{equation}
This can attains the Carnot efficiency for time-reversal symmetric case $x=1$. The $\eta_{max}$ is plotted (dotted curve) as a function of the asymmetric parameter $x$ is shown in Fig. \ref{fig:fig2}. Maximizing the power output (Eq. (\ref{1ab})) with respect to $X_{\rho}$, keeping  $X_{q}$ fixed, we get \cite{van}
\begin{equation}\label{1ad}
X_{\rho}^{P_{max}} = \frac{X_{\rho}^{stop}}{2}; \;X_{\rho}^{stop}\equiv-\frac{L_{\rho q}}{L_{\rho\rho}}X_q.
\end{equation}
The system halts for the external force $F=X_{\rho}^{stop} T$, i.e., $J_\rho=0$ \cite{van}. Substituting $X_{\rho}^{P_{max}}$ in  Eqs. (\ref{1ab}) and (\ref{1ac}), we get, respectively, the maximum power output and the efficiency at maximum power output as \cite{ben}
\begin{equation}\label{1ae}
P_{max}=\frac{\eta_C}{4}\frac{L_{\rho q}^{2}}{L_{\rho\rho}} X_q,
\end{equation}
\begin{equation}\label{1af}
\eta(P_{max})=\frac{\eta_C}{2} \frac{1}{2\frac{L_{\rho\rho} L_{qq}}{L_{\rho q}^{2}}-\frac{L_{q\rho}}{L_{\rho q}}}.
\end{equation}
Using Eqs. (\ref{1z}) and (\ref{1aa}), we can write the efficiency at maximum power output as \cite{ben}
\begin{equation}\label{1ag}
\eta(P_{max})= \eta_C \frac{xy}{4+2y}.
\end{equation} 
For a given asymmetry parameter $x$, the efficiency at maximum power output attains its maximum when $4y=h(x)$, which is given by \cite{bra1,bra2}
\begin{equation}\label{1aj}
\eta(P_{max})= \eta_C \frac{x^{2}}{4x^{2}-6x+4}.
\end{equation}
Fig. \ref{fig:fig2} shows the $\eta(P_{max})$ is plotted (dashed curve) as a function of the asymmetry parameter $x$. The $\eta(P_{max})$ attains $\eta_C/2$ for time-reversal symmetric case and reaches its maximum of $4\eta_C/7$ at $x=4/3$.
\subsection{The efficiency under maximum $\dot{\Omega}$ criterion}
	Now, we analyze a thermoelectric heat engine under the maximum $\dot{\Omega}$ criterion which accounts for both useful energy and losses. The target function is given as $\dot{\Omega} = 2P-\eta_{max} J_q$ \cite{fnote2}. Using Eqs. (\ref{1o2}) and (\ref{1ab}), we get
\begin{eqnarray}\label{1ao}
\dot{\Omega}& =& -2T J_{\rho}  X_{\rho}-\eta_{max} J_q\\ \nonumber
&=&- 2TL_{\rho\rho}X_{\rho}^{2}-(2L_{\rho q}\eta_C +L_{q\rho}\eta_{max})X_{\rho}- L_{qq}\eta_{max} X_q.
\end{eqnarray}
Maximizing the above $\dot{\Omega}$ criterion with respect to $X_{\rho}$, keeping $X_q$ fixed, we get
\begin{equation}\label{1ap}
X_{\rho}^{\dot{\Omega}_{max}}=-\frac{1}{2L_{\rho\rho}}\left(L_{\rho q}+ \frac{L_{q \rho}}{2}\frac{\eta_{max}}{\eta_C}\right)X_q.
\end{equation}
Substituting Eq. (\ref{1ap}) in Eqs. (\ref{1ab}) and (\ref{1ac}), we get, respectively, the power output and the efficiency under the maximum $\dot{\Omega}$ criterion as
\begin{equation}\label{1aq}
P(\dot{\Omega}_{max})=P_{max}\left(1- \frac{L_{q\rho}^{2}}{4L_{\rho q}^{2}}\frac{\eta_{max}^2}{\eta_C^{2}}\right),
\end{equation}
\begin{equation}\label{1ar}
\eta(\dot{\Omega}_{max})= \frac{\frac{\eta_C}{2}-\frac{L_{q\rho}^{2}}{8L_{\rho q}^{2}}\frac{\eta_{max}^2}{\eta_C}}{\frac{2L_{\rho\rho} L_{qq}}{L_{\rho q}^{2}}-\frac{L_{q \rho}}{L_{\rho q}}-\frac{L_{q\rho}^{2}}{2L_{\rho q}^{2}}\frac{\eta_{max}}{\eta_C}}.
\end{equation}
Using Eqs. (\ref{1z}), (\ref{1aa}) and (\ref{1ae}), we can write Eqs. (\ref{1aq}) and (\ref{1ar}) as
\begin{equation}\label{1as}
P(\dot{\Omega}_{max}) = P_{max}\left(1-\frac{1}{4x^{2}}\frac{\eta_{max}^2}{\eta_C^{2}}\right),
\end{equation}
\begin{equation}\label{1at}
\eta(\dot{\Omega}_{max})=\frac{(4\eta_C^{2}x^{2}-\eta_{max}^{2})y}{4[2\eta_C(2+y)x-\eta_{max}y]}.
\end{equation}
For a given asymmetry parameter $x$, the $\eta(\dot{\Omega}_{max})$ attains its maximum when $4y=h(x)$, which is given by
\begin{equation}\label{1au}
\eta(\dot{\Omega}_{max})=\frac{4\eta_C^{2}x^{2}-\eta_{max}^{2}}{4[2\eta_C(2x^{2}-3x+2)-\eta_{max}]}.
\end{equation}
Substituting Eq. (\ref{1am}) in Eqs. (\ref{1as}) and (\ref{1au}), we get the power output and the efficiency under the maximum $\dot{\Omega}$ criterion, respectively, as
\begin{equation}\label{1ca}
P(\dot{\Omega}_{max})=P_{max}\frac{6x^{2}-9x+6+10\sqrt{x^{2}-x+1}|x-1|}{4[2x^{2}-3x+2+2\sqrt{x^{2}-x+1}|x-1|]},
\end{equation}
\begin{equation}\label{1av}
\eta(\dot{\Omega}_{max})=\frac{3}{4}\eta_C x^{2}\frac{2x^{2}-3x+2+(10/3)\sqrt{x^{2}-x+1}|x-1|}{8x^{4}-24x^{3}+33x^{2}-24x+8+(8x^{2}-12x+8)\sqrt{x^{2}-x+1}|x-1|}.
\end{equation}  
For time-reversal symmetric case $x=1$, we get $\eta(\dot{\Omega}_{max})=3\eta_C/4$, which is the lower bound obtained for the asymmetrical dissipation limits of both the low dissipation heat engines and the minimally nonlinear irreversible heat engines under the tight-coupling condition (i.e., without heat leakage between the system and the reservoirs \cite{ked}) \cite{det2,lon}. 
	
	When a thermoelectric heat engine with broken time-reversal symmetry is working under the maximum $\dot{\Omega}$ criterion, the efficiency lies between the maximum efficiency and the efficiency at maximum power output as shown in Fig. \ref{fig:fig2}. When the asymmetry parameter $x$ is slightly larger than 1, the efficiency under the maximum $\dot{\Omega}$ criterion overcomes $3\eta_C/4$ and attains maximum of $0.75661\eta_C$ at $x=1.05764$. As compared to the efficiency at maximum power output, the $\eta(\dot{\Omega}_{max})$ increases sharply towards its maximum value from both sides of the asymmetry parameter near its symmetric value. This shows that a thermoelectric heat engine working under the maximum $\dot{\Omega}$ criterion provides a  significantly better performance than a device working at maximum power output. When $|x| \rightarrow \infty$, both the $\eta(\dot{\Omega}_{max})$ and $\eta(P_{max})$ asymptotically approaches to $\eta_C/4$. The target function under the maximum $\dot{\Omega}$ criterion is given by
\begin{equation}\label{1aw}
\dot{\Omega}_{max}=\frac{\eta_C^{2}}{2TL_{\rho\rho}}\left[\left(L_{\rho q}-\frac{L_{q\rho}}{2}\frac{\eta_{max}}{\eta_C}\right)^{2}-2\textbf{Det} \, \textbf{L}\frac{\eta_{max}}{\eta_C}\right].
\end{equation}
Following Ref. \cite{ben2}, we now analyze the nature of the efficiency bound on power output. The normalized power output and the normalized efficiency are defined as
\begin{equation}\label{1ax}
\bar{P}\equiv\frac{P}{P_{max}};\quad\bar{\eta}\equiv\frac{\eta}{\eta_C}.
\end{equation}
We can write the normalized power output as a function of $X_{\rho}$ with $X_q=-L_{\rho\rho}X_{\rho}^{stop}/L_{\rho q}$ as
\begin{equation}\label{1ay}
\bar{P}=4k(1-k),
\end{equation}
where $k=X_{\rho}/X_{\rho}^{stop}$. From Eq. (\ref{1ay}), we get 
\begin{equation}\label{1az}
k=\frac{1\pm \sqrt{\strut 1-\bar{P}}}{2}.
\end{equation}
In above equation, the positive sign gives $k\geq1/2$ and the negative sign gives $k\leq1/2$. Using Eqs. (\ref{1ac}) and (\ref{1az}), we get the normalized efficiency as
\begin{equation}\label{1ba}
\bar{\eta}_{\pm}=\bar{P}\frac{xy}{2\left[2+\left(1\mp\sqrt{\strut1-\bar{P}}\right)y\right]}.
\end{equation}
Where $\bar{\eta}_{+}$ denote the normalized efficiency for $k\geq1/2$ and $\bar{\eta}_{-}$ for $k\leq1/2$. For a given asymmetry parameter $x$, the $\bar{\eta}_{\pm}$ attains its maximum when $4y=h(x)$, which is given by
\begin{equation}\label{1bb}
\bar{\eta}_{\pm}=\bar{P}\frac{x^{2}}{2\left(2x^{2}-3x+2\mp x\sqrt{\strut1-\bar{P}}\right)}.
\end{equation}
\begin{figure}[h]
\centering
    \includegraphics[width=0.50\linewidth]{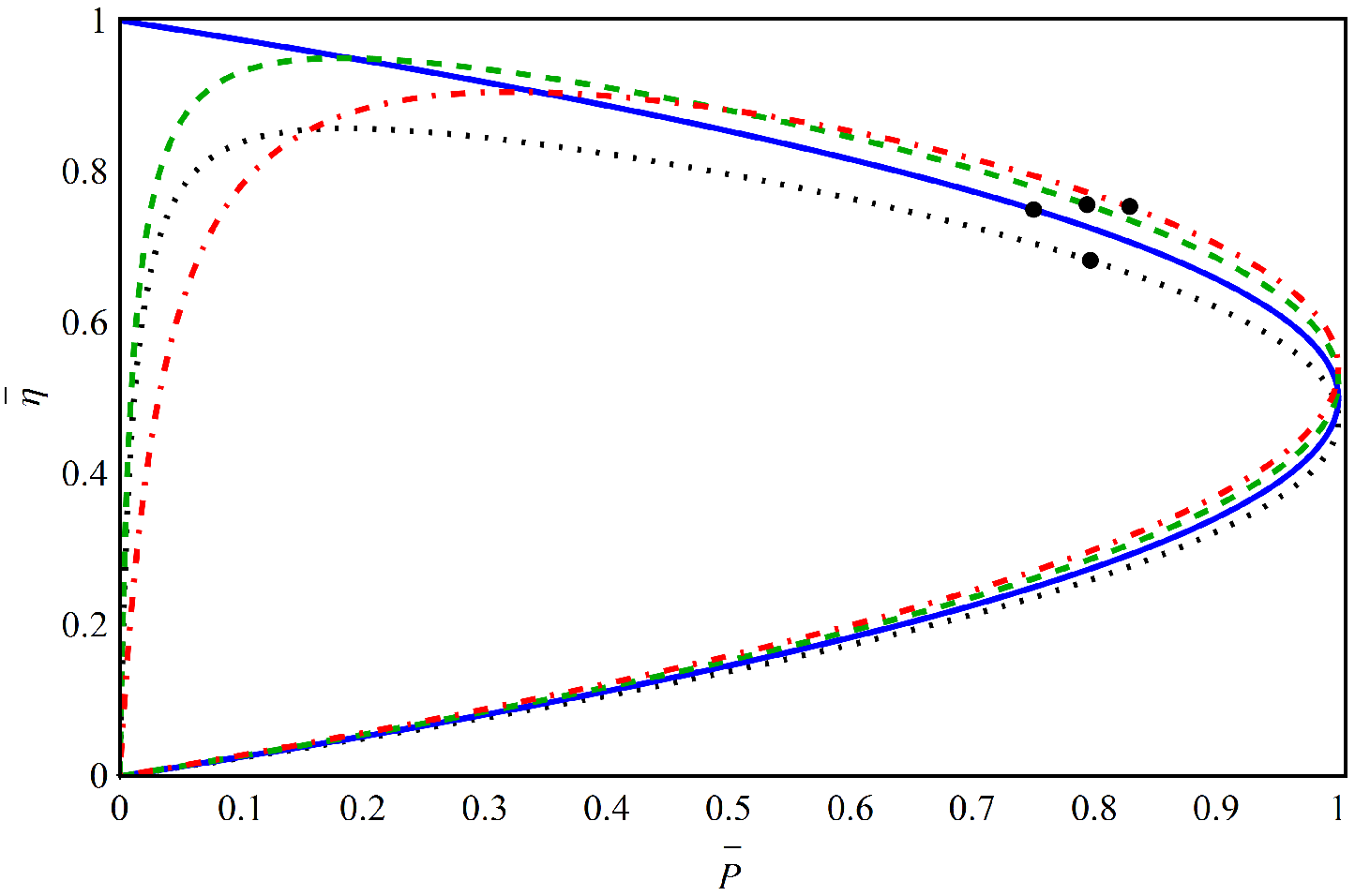}
   \label{fig:Ng1} 
   \includegraphics[width=0.50\linewidth]{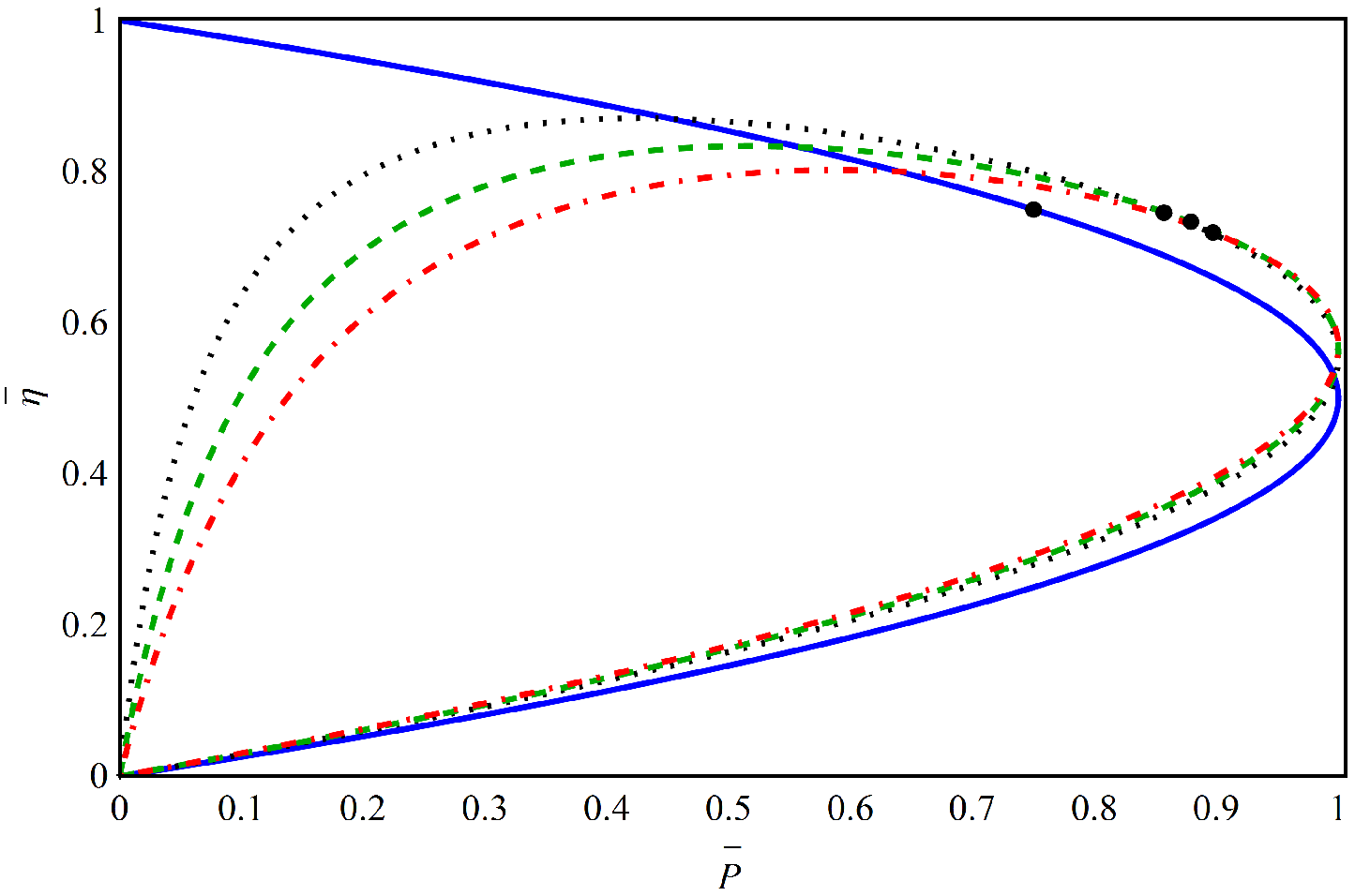}
    \label{fig:Ng2}
\caption{The normalized efficiency (Eq. (\ref{1bb})) is plotted as a function of the normalized power output ($\bar{P}$) for various values of the asymmetry parameter $x$. In both figures, the upper part of the curves belong to $\bar{\eta}_{+}$ and the lower part  belong to $\bar{\eta}_{-}$. In both figures the solid curve represent for the asymmetry parameter $x=1$. The figure (top) is plotted for the asymmetry parameters, $x=0.95$ (dotted curve), $x=1.05$ (dashed curve), $x=1.1$ (dot-dashed curve). The figure (bot) is plotted for the asymmetry parameters, $x=1.15$ (dotted curve), $x=1.2$ (dashed curve), $x=1.25$ (dot-dashed curve). The dots in the figures are the normalized efficiency corresponding to the normalized power under the maximum $\dot{\Omega}$ criterion for different values of $x$.}
\label{fig:fig3} 
\end{figure}
Fig. \ref{fig:fig3} shows the $\bar{\eta}_{\pm}$ is plotted as a function of the normalized power for various values of the asymmetry parameter $x$. For $x=1$, the normalized efficiency increases towards its maximum value when a thermoelectric engine working at a lower power than the maximum. When the asymmetry parameter $x$ is slightly larger than $1$, the normalized efficiency exceeds its time-reversal symmetric case $x=1$ over a wider range of the normalized power output. The dots in the figures are the normalized efficiency (Eq. (\ref{1av})) corresponding to the normalized power output (Eq. (\ref{1ca})) under the maximum $\dot{\Omega}$ criterion for different values of the asymmetric parameter $x$. Which represents the best compromise between the power output and efficiency and it shows that for a small decrease of power output enhances the efficiency considerably. Within a small range of finite power output, Fig. \ref{fig:fig3} (top) shows that the $\eta(\dot{\Omega}_{max})$ is higher than its time-reversal symmetric case for the asymmetric parameter $x$ is slightly larger than one. The dots in Fig. \ref{fig:fig3} (bot) shows that the normalized efficiency under the maximum $\dot{\Omega}$ criterion moves towards the maximum normalized power output for a larger asymmetry. When $|x|\rightarrow\infty$, $\bar{\eta}_{\pm}\rightarrow\bar{P}/4$ and hence $\bar{\eta}_{\pm}$ can attain $1/4$ for maximum power output. 
\section{A Thermoelectric Refrigerator}
	The system work as a refrigerator, for $J_q<0$ and $P<0$ (i.e., $J_{\rho}<0$) \cite{ben,ben2}. In a steady-state, the coefficient of performance of a thermoelectric refrigerator is defined as the ratio of the heat current extracted from the cold reservoir and the consumed power \cite{bra1}, which is given by
\begin{equation}\label{1bc}
\eta^{r}= \frac{-J_q}{TJ_{\rho}X_{\rho}}=\frac{-(L_{q\rho}X_{\rho}+L_{qq}X_q)}{T(L_{\rho\rho}X_{\rho}^{2}+L_{\rho q}X_{\rho}X_q)}.
\end{equation}
\subsection{The coefficient of performance at maximum cooling load}
To find the maximum coefficient of performance, we maximize Eq. (\ref{1bc}) with respect to $X_{\rho}$ for fixed $X_q$ with the condition $J_q<0$ and $P<0$, we get \cite{ben}
\begin{equation}\label{1bi}
X_{\rho}^{\eta_{max}^{r}}=\frac{L_{qq}}{L_{q\rho}}\left(-1-\sqrt{\frac{\textbf{Det} \, \textbf{L}}{L_{\rho\rho}L_{qq}}}\,\right)X_q,
\end{equation}
substituting $X_{\rho}^{\eta_{max}^{r}}$ in Eq. (\ref{1bc}), we get the maximum coefficient of performance as \cite{ben}
\begin{equation}\label{1bj}
\eta_{max}^{r}= \eta^{r}_C\frac{1}{x}\frac{\sqrt{y+1}-1}{\sqrt{y+1}+1},
\end{equation}
where the Carnot coefficient of performance $\eta^{r}_C=1/(T X_q)$. The $\eta_{max}^{r}$ attains its maximum when $4y=h(x)$, which is given by \cite{bra2}
\begin{equation}\label{1bk}
\eta_{max}^{r}=\eta_C^{r}\frac{1}{x}\frac{\sqrt{x^{2}-x+1}-|x-1|}{\sqrt{x^{2}-x+1}+|x-1|}.
\end{equation}
For time-reversal symmetric case $x=1$, the $\eta_{max}^{r}$ attains the Carnot coefficient of performance. The $\eta_{max}^{r}$ is  plotted (dotted curve) as a function of the asymmetric parameter $x$ is shown in Fig. \ref{fig:fig4}. The maximum coefficient of performance ($\eta^{r}_{max}$) sharply increases towards $x=1$ on either side. 

The coefficient of performance at maximum cooling power were studied in finite-time thermodynamics for specific models of refrigerators \cite{agra,ape}. However, the COP at maximum cooling power and the COP at maximum $\chi$ criterion is not yet accomplished in the general setting of the linear irreversible thermodynamics \cite{ape,joh,shen}. In such a case, de Cisneros \textit{et al}. showed that the COP at maximum $J_q X_q$ for a refrigerator is equivalent to the Curzon-Alhborn efficiency at maximum power of a heat engine \cite{cisn}. Hence, in our analysis we use the  quantity $P^{r}\equiv J_q X_q$ called as the cooling load for optimization and obtain the COP of a thermoelectric refrigerator at maximum cooling load. By using Eq. (\ref{1o2}) the magnitude of cooling load is given as \cite{cisn,din}.
\begin{equation}\label{1bd}
P^{r}= L_{q\rho}X_{\rho} X_q+L_{qq} X_q^{2}.
\end{equation}
Maximizing the above cooling load with respect to $X_q$, by keeping $X_{\rho}$ fixed, we get 
\begin{equation}\label{1be}
X_q^{P^{r}_{max}} = \frac{X_q^{stop}}{2}; \;X_q^{stop}=-\frac{L_{q\rho}}{L_{qq}} X_{\rho}.
\end{equation}
The input heat flux $J_q$ becomes zero at $X_q=X_q^{stop}$, then a refrigerator halts. Substituting Eq. (\ref{1be}) in Eqs. (\ref{1bc}) and (\ref{1bd}), we get, respectively, the magnitude of maximum cooling load and the COP at maximum cooling load as
\begin{equation}\label{1mm}
P^{r}_{max}=\frac{L_{q\rho}^{2}}{4L_{qq}}X_{\rho}^{2}.
\end{equation}
\begin{equation}\label{1bf}
\eta^{r}(P^{r}_{max})=\frac{\eta^{r}_C}{2} \frac{1}{\frac{2 L_{\rho\rho}L_{qq}}{L_{q\rho}^{2}} -\frac{L_{\rho q}}{L_{q\rho}}}.
\end{equation}
Using Eqs. (\ref{1z}) and (\ref{1aa}), we can write the Eq. (\ref{1bf}) as
\begin{equation}\label{1bg}
\eta^{r}(P^{r}_{max})=\frac{\eta^{r}_C}{x}\frac{y}{4+2y}.
\end{equation}
For a given asymmetry parameter $x$, the $\eta^{r}(P^{r}_{max})$ attains its maximum when $4y=h(x)$, which is given by
\begin{equation}\label{1bh}
\eta^{r}(P^{r}_{max})=\eta^{r}_C\frac{1}{4x^{2}-6x+4}.
\end{equation}
Fig. \ref{fig:fig4} shows the $\eta^{r}(P^{r}_{max})$ is plotted (dashed curve) as a function of the asymmetric parameter $x$. The $\eta^{r}(P^{r}_{max})$ attains the Curzon-Ahlborn coefficient of performance $\eta_{CA}^{r}=\eta_C^{r}/2$ \cite{cisn} for time-reversal symmetric case and overcomes in a small range of $x$ less than $1$ with attaining the maximum of $4\eta_C^{r}/7$ at $x=3/4$. The $\eta^{r}(P^{r}_{max})$ approaches zero, when $|x| \rightarrow \infty$. 
\subsection{The coefficient of performance under $\dot{\Omega}$ criterion}
	Now, we analyze a thermoelectric refrigerator under maximum $\dot{\Omega}$ criterion. For refrigerator, the target function is given as $\dot{\Omega}= -2 J_q-\eta_{max}^{r}P$. The consumed power $P=T J_{\rho}  X_{\rho}$. Using Eqs. (\ref{1o}) and (\ref{1o2}), we get
\begin{eqnarray}\label{1bm}
\dot{\Omega} &=& -2 J_q-\eta_{max}^{r}T J_{\rho}  X_{\rho}\\ \nonumber
&=& -TL_{\rho\rho}\eta_{max}^{r}X_{\rho}^{2}-\left(2L_{q\rho}+L_{\rho q}\frac{\eta_{max}^{r}}{\eta_C^{r}}\right) X_{\rho}-2   L_{qq} X_q.
\end{eqnarray}	
Maximizing the above $\dot{\Omega}$ criterion with respect to $X_{\rho}$, keeping $X_q$ fixed, we get
\begin{equation}\label{1bn}
X_{\rho}^{\dot{\Omega}_{max}}=-\frac{1}{L_{\rho\rho}}\left(\frac{L_{\rho q}}{2}+L_{q\rho}\frac{\eta_C^{r}}{\eta_{max}^{r}}\right) X_q.
\end{equation}
Substituting Eq. (\ref{1bn}) in Eqs. (\ref{1bc}) and (\ref{1bd}), we get the cooling load and the COP under the maximum $\dot{\Omega}$ criterion, respectively, as
\begin{equation}\label{1bo}
P^{r}(\dot{\Omega}_{max})= \frac{1}{2L_{\rho\rho}}\left(2\textbf{Det} \, \textbf{L}+L_{\rho q}L_{q\rho}-2L_{q\rho}^{2}\frac{\eta_C^{r}}{\eta_{max}^{r}}\right)X_q^{2},
\end{equation}
\begin{equation}\label{1bp}
\eta^{r}(\dot{\Omega}_{max})= \eta^{r}_C \frac{2\frac{L_{\rho\rho}L_{qq}}{L_{\rho q}^{2}}- \frac{L_{q\rho}}{ L_{\rho q}} -2\frac{L_{q\rho}^{2}}{L_{\rho q}^{2}} \frac{\eta_C^{r}}{\eta_{max}^{r}}}{\frac{1}{2}-2\frac{L_{q\rho}^{2}}{L_{\rho q}^{2}}\left(\frac{\eta_C^{r}}{\eta_{max}^{r}}\right)^{2}}.
\end{equation}
Using Eqs. (\ref{1z}) and (\ref{1aa}), we get Eq. (\ref{1bp}) as
\begin{equation}\label{1bq}
\eta^{r}(\dot{\Omega}_{max})=\eta^{r}_C\frac{2[x(2+y)\eta_{max}^{r}-2y\eta_{C}^{r}]\eta_{max}^{r}}{y(x^{2}\eta_{max}^{r2} -4\eta_C^{r2})}.
\end{equation}
For a given value of the asymmetry parameter $x$, the $\eta^{r}(\dot{\Omega}_{max})$ attains its maximum when $4y=h(x)$, which is given by
\begin{equation}\label{1br}
\eta^{r}(\dot{\Omega}_{max})= \eta_C^{r}\frac{2[(2x^{2}-3x+2)\eta_{max}^{r}-2\eta_C^{r}]\eta_{max}^{r}}{ x^{2}\eta_{max}^{r2} -4\eta_C^{r2}}.
\end{equation}
Substituting Eq. (\ref{1bk}) in Eq. (\ref{1br}), we get
\begin{equation}\label{1bs}
\eta^{r}(\dot{\Omega}_{max})=\frac{2}{3}\frac{\eta_C^{r}}{x^{2}}\frac{(4x^{2}-6x+4)\sqrt{x^{2}-x+1}|x-1|-4x^{4}+12x^{3}-15x^{2}+12x-4}{2x^{2}-3x+2+(10/3)\sqrt{x^{2}-x+1}|x-1|}.
\end{equation}
For time-reversal symmetric case $x=1$, we get $\eta^{r}(\dot{\Omega}_{max})=2\eta_C^{r}/3$, which is the lower bound on the coefficient of performance under the maximum $\dot{\Omega}$ criterion obtained for the asymmetrical dissipation limits of both the low-dissipation refrigerators and the minimally nonlinear irreversible refrigerators under the tight-coupling condition \cite{lon,det2}.
\begin{figure}[h]
	\centering
	\includegraphics[width=0.50\textwidth]{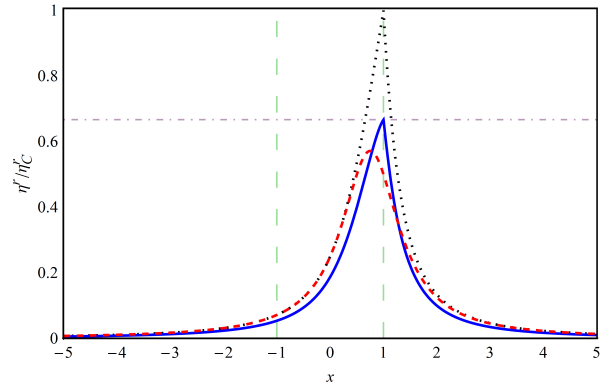}
	\caption{\label{fig:fig4}The normalized coefficient of performance ($ \eta^{r}/\eta^{r}_C $) is plotted as a function of the asymmetry parameter $x$. In the dotted curve $\eta^{r}=\eta^{r}_{max}$, solid curve for $ \eta^{r}=\eta^{r}(\dot{\Omega}_{max})$ and the dashed curve for $\eta^{r}=\eta^{r}(P^{r}_{max})$. The vertical dashed line represents $|x|=1$ and a horizontal dot-dashed line indicates $\eta^{r}=3\eta_C^{r}/2$.}
\end{figure}		 

The $\eta^{r}(\dot{\Omega}_{max})$ plotted (solid curve) as a function of the asymmetric parameter $x$ is shown in Fig.  \ref{fig:fig4}. Since there is no proper optimization criterion available in literature for optimizing only with $X_{\rho}$ \cite{det,tay}, we have maximized the cooling load with respect to $X_q$ and the $\dot{\Omega}$ criterion with respect to $X_{\rho}$. This restricts us to strictly compare the COP under the maximum $\dot{\Omega}$ criterion with the COP at maximum cooling load (see Eqs. (\ref{1mm}) and (\ref{1bo})).  In such a case, we find that when a thermoelectric refrigerator working under the maximum $\dot{\Omega}$ criterion, its coefficient of performance ($\eta^{r}(\dot{\Omega}_{max})$) is larger than the coefficient of performance at maximum cooling load only for a small range of the asymmetry parameter in the vicinity of $x\approx 1$. For $|x|$ much larger than $1$, the coefficient of performance at maximum cooling load is higher than the coefficient of performance under the maximum $\dot{\Omega}$ criterion. The target function under the maximum $\dot{\Omega}$ criterion is given by
\begin{equation}\label{1bt}
\dot{\Omega}_{max}= \frac{\eta_{max}^{r}}{T L_{\rho\rho}\eta_C^{r2}}\left[\left(\frac{L_{\rho q}}{2}-L_{q\rho}\frac{\eta_{max}^{r}}{\eta_C^{r}}\right)^{2}-2 \textbf{Det} \, \textbf{L} \frac{\eta_C^{r}}{\eta_{max}^{r}} \right].
\end{equation}
Next, we analyze the nature of the COP bound on cooling load. The normalized cooling load and the normalized COP is defined as
\begin{equation}\label{1bu}
\bar{P^{r}}\equiv\frac{P^{r}}{P^{r}_{max}};\quad\bar{\eta^{r}}\equiv\frac{\eta^{r}}{\eta^{r}_C}.
\end{equation}
We can write the normalized cooling load as a function of $X_q$ with $X_{\rho}=-L_{qq}X_q^{stop}/L_{q\rho}$ as
\begin{equation}\label{1bv}
\bar{P^{r}}=4l(1-l),
\end{equation}
where $l=X_q/X_q^{stop}$. From Eq. (\ref{1bv}), we get
\begin{equation}\label{1bw}
l=\frac{1\pm \sqrt{\strut1-\bar{P^{r}}}}{2}.
\end{equation}
In the above equation, the positive sign gives $l\geq1/2$ and the negative sign gives $l\leq1/2$. Substituting Eq. (\ref{1bw}) in  Eq. (\ref{1bc}), we get the normalized COP as
\begin{equation}\label{1bx}
\bar{\eta^{r}}_{\pm}=\bar{P^{r}}\frac{1}{x}\frac{y}{2\left[2+\left(1\mp\sqrt{\strut1-\bar{P^{r}}}\right)y\right]}.
\end{equation}
Where $\bar{\eta^{r}}_{+}$ denote the normalized coefficient of performance for $l\geq1/2$ and the $\bar{\eta^{r}}_{-}$ for $l\leq1/2$. The $\bar{\eta^{r}}_{\pm}$ attains its maximum value for a given $x$ when $4y=h(x)$, which is given by
\begin{equation}\label{1by}
\bar{\eta^{r}}_{\pm}=\bar{P^{r}}\frac{1}{2 \left(2x^{2}-3x+2\mp x\sqrt{\strut1-\bar{P^{r}}}\right)}.
\end{equation}
\begin{figure}[h]
\centering
\includegraphics[width=0.48\textwidth]{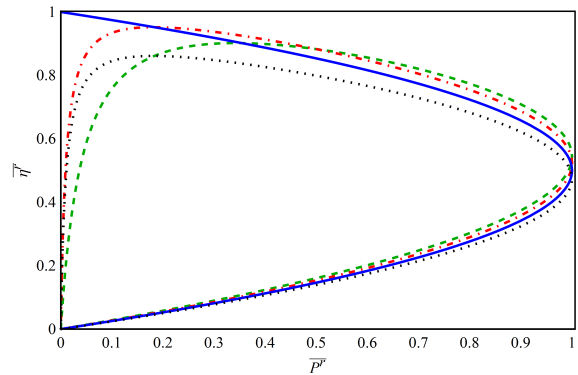}
\caption{ The normalized coefficient of performance (Eq. (\ref{1by})) is plotted as a function of the normalized cooling load for the various values of the asymmetry parameter $x$. In all curves, the upper part belong to $\bar{\eta^{r}}_{+}$ and the lower part belong to $\bar{\eta^{r}}_{-}$. The figure is plotted for the asymmetry parameters, $x=0.90$ (dashed curve), $x=0.95$ (dot-dashed curve), $x=1$ (solid curve), $x=1.05$ (dotted curve).}
\label{fig:fig5}
\end{figure}

As previously mentioned, the use of different optimization parameters for cooling load and $\dot{\Omega}$ criterion of a refrigerator restricts us to compare the coefficient of performance at maximum cooling load with the coefficient of performance under the maximum $\dot{\Omega}$ criterion (see Eqs. (\ref{1mm}) and (\ref{1bo})). However, $\bar{\eta^{r}}_{\pm}$ plotted as a function of the normalized cooling load for various values of the asymmetry parameter $x$ is shown in Fig. \ref{fig:fig5}. It indicates that the coefficient of performance of a thermoelectric refrigerator can attain the Carnot coefficient of performance with zero cooling load only for the time-reversal symmetric case $x=1$. The normalized coefficient of performance has larger values when it working at a lower cooling load than the maximum. When the asymmetry parameter $x$ is slightly lower than $1$, the normalized coefficient of performance exceeds its time-reversal symmetric case over a wider range of the normalized cooling load. When the asymmetric parameter $|x|\rightarrow\infty$, $\bar{\eta^{r}}_{\pm}\rightarrow0$, which means the device no longer cools the system.

\section{Conclusion}

	We incorporated the generalized framework of LIT and studied the performance of a thermoelectric device, namely the heat engines efficiency and the refrigerators coefficient of performance, with broken time-reversal symmetry under the maximum $\dot{\Omega}$ criterion. We showed that when a thermoelectric heat engine is working under the maximum $\dot{\Omega}$ criterion its  efficiency ($\eta(\dot{\Omega}_{max})$) increases significantly as compared to the efficiency at maximum power output ($\eta(P_{max})$). For time-reversal symmetric case $x=1$, we get the lower bound $\eta(\dot{\Omega}_{max})=3\eta_C/4$ obtained for the low-dissipation Carnot heat engines and the minimally nonlinear irreversible heat engines with the asymmetrical dissipation limits and the tight-coupling condition. In a narrow range of the asymmetry parameter $x$ slightly larger than $1$, a thermoelectric heat engine overcomes $3\eta_C/4$. When $|x|\rightarrow \infty$, both the $\eta(\dot{\Omega}_{max})$ and $\eta(P_{max})$ asymptotically approaches $\eta_C/4$.
	
	The coefficient of performance of a thermoelectric refrigerator under the maximum $\dot{\Omega}$ criterion ($\eta^{r}(\dot{\Omega}_{max})$) has a larger value than the coefficient of performance at maximum cooling load ($\eta^{r}(P^{r}_{max})$) only for a narrow range of the asymmetry parameter in the vicinity of $x=1$. For time-reversal symmetric case $x=1$, we get the lower bound $\eta^{r}(\dot{\Omega}_{max})=2\eta_C^{r}/3$ obtained for the low-dissipation Carnot refrigerators and the minimally nonlinear irreversible refrigerators with the asymmetrical dissipation limits and the tight-coupling condition. When $|x|\rightarrow \infty$, both the $\eta^{r}(\dot{\Omega}_{max})$ and $\eta^{r}(P^{r}_{max})$ asymptotically approaches zero.
	
	A slight increase (decrease) in asymmetry parameter from its time-reversal symmetric case ($x=1$) enhances the efficiency (coefficient of performance) of a thermoelectric heat engine (refrigerator) with finite power output (cooling load). Our results may helpful to design and operate a real thermoelectric device with broken time-reversal symmetry for a productive outcome. Apertet \textit{et al}. studied an autonomous thermoelectric generator by extending the local Onsager relations to a global scale and showed that it naturally includes the Joule heat dissipation in input and output heat fluxes \cite{ape2}, which we will consider in our future study. The present study can also be extended to a thermoelectric heat devices beyond the linear response regime. Further studies of a multi-terminal thermoelectric device \cite{bra2}, quantum thermoelectric device \cite{whi} and the microscopic thermoelectric device \cite{yam2}, under the maximum $\Omega$ criterion will be helpful for efficient energy harvesting.

\begin{center}
\textbf{ACKNOWLEDGEMENT}
\end{center}
One of the author I.I would like to thank G. Phanindra Narayan, K. R. S. Preethi Meher, R. Raveendraprathap, H. Saveetha, Joseph Prabagar, K. Nilavarasi, A. Arul Anne Elden and Hebrew Benhur Crispin for useful discussions and valuable suggestions.


\begin{thebibliography}{100}
\bibitem{fnote1}
The discussion about the difference between the chemical potential and the electrochemical potential can be found in \cite{casa,aper}. Since the electrochemical potential becomes identical to the chemical potential when the reservoir is electrically neutral we use the chemical potential instead of the electrochemical potential, which is the conventional usage in the study of a thermoelectric device \cite{ben,sai1,bra1,casa}.
\bibitem{fnote2}
In cyclic heat engine, the maximum efficiency for given input heat $Q_q$ is considered as $\eta_{max}=W_{max}/Q_q$ \cite{her}.  Since we study a steady state heat engine, we use the definition of maximum efficiency for given input heat flux $\dot{Q_q}$ as $\eta_{max}=\left(\dot{W}/\dot{Q_q}\right)_{max}$ \cite{ben}.
\end{thebibliography}
\end{document}